\documentclass[10pt,letterpaper,journal]{IEEEtran}
\usepackage[cmex10]{amsmath}
\usepackage{amsfonts,amsthm,bm}
\usepackage{array,cite,eqparbox,dcolumn}
\usepackage{graphicx}
\usepackage{subfig}
\usepackage{graphicx}
\usepackage{dcolumn}
\usepackage{bm}
\usepackage{amsthm}
\newtheorem{theorem}{Theorem}
\newtheorem{conjecture}{Conjecture}
\newtheorem{result}{Result}
\graphicspath{{figures/pdf/}}
\DeclareGraphicsExtensions{.pdf}
\def\vec#1{{\bm #1}}

\def\ket#1{| #1 \rangle}
\def\bra#1{\langle #1 |}
\def\ip#1#2{\langle #1 | #2 \rangle}

\def\ZZ{\mathbb{Z}}

\def\diag{\operatorname{diag}}

\def\dim{\operatorname{dim}}

\def\Tr{\operatorname{Tr}}
\def\gcd{\operatorname{gcd}}

\def\H{\mathcal{H}}

\def\ONE{\mathbb{I}}

\def\so{\mathfrak{so}}
\def\sp{\mathfrak{sp}}
\def\su{\mathfrak{su}}
\def\uu{\mathfrak{u}}
\def\u{\mathfrak{u}}

\def\LL{\mathfrak{L}}

\def\up{{\uparrow}}
\def\down{{\downarrow}}
\begin{document}
\title{Symmetry \& Controllability for Spin Networks with a Single-Node
Control}


\author{\IEEEauthorblockN{Xiaoting Wang\IEEEauthorrefmark{1}, 
        Peter Pemberton-Ross\IEEEauthorrefmark{1} and
	S G Schirmer\IEEEauthorrefmark{1}\\
	\IEEEauthorblockA{\IEEEauthorrefmark{1}
                  Dept of Applied Mathematics \& Theoretical Physics,
                  University of Cambridge,\\
                  Wilberforce Road, Cambridge, CB3 0WA, UK\\
           Email: x.wang@damtp.cam.ac.uk, pjp32@cam.ac.uk, sgs29@cam.ac.uk}}}

\date{\today}

\maketitle

\begin{abstract}
We consider the relation of symmetries and subspace controllability for
spin-$\tfrac{1}{2}$ networks with XXZ couplings, subject to perturbation
of a single node by a local potential ($Z$-control).  The Hamiltonians
for such networks decompose into excitation subspaces.  Focusing on the
single excitation subspace, it is shown for single-node $Z$-controls
that external symmetries are characterized by eigenstates of $H_0$ which
have zero overlap with the control node, and there are no internal
symmetries. It is further shown that there are symmetries which persist
even in the presence of random perturbations.  For XXZ chains with
uniform coupling strengths, a characterization of all possible
symmetries is given which shows a strong dependence on the position of
the node we control.  We then show for Heisenberg and XX chains with
uniform coupling strength subject to single-node $Z$-control that the
lack of symmetry is both necessary and sufficient for subspace
controllability.  Finally, the latter approach is generalized to
establish controllability results for simple branched networks.
\end{abstract}



\section{Introduction}

In the study of quantum communication and quantum computation, spin
chains and general spin networks are simple but very useful models to
approximate physical
systems~\cite{Benjamin-Bose,Lloyd-Landahl,Kay-Ericsson,Vollbrecht-Cirac},
including molecules in NMR experiments~\cite{Chuang} and ultracold atoms
in optical lattices~\cite{Cirac}.  In the latter case it has been
demonstrated experimentally that single and two-qubit gates can be
implemented by controlling the external magnetic field, thus realizing
universal quantum computation (QC)~\cite{universal-book,Div,Barenco}.
Universality is the capability of generating all unitary computational
gates, which in the language of control theory is equivalent to system
controllability, i.e., the capability to generate the Lie algebra of the
unitary (or special unitary) group using both the system Hamiltonian and
the available control Hamiltonians~\cite{Jurdjevic,D'Alessandro-book}.
Controllability is an important criterion for assessing a system's
capability for quantum information processing and many other
applications of quantum control~\cite{Altafini_1,FSS-01,
Albertini_Dalessandro_1, Albertini_Dalessandro_2,SLS-02,SPS-03,SPS-05,
Albertini_Dalessandro_3}.  In principle, controllability can be verified
by explicitly computing the dimension of the Lie algebra, but such
computations quickly become intractable as the dimension of the system
increases.  Therefore, alternative controllability conditions are
needed, and many of these have been studied and derived for spin
networks of various couplings and
topologies~\cite{Albertini_Dalessandro_1,Khaneja,Burgarth-Bose}.  Most
of the early results characterize conditions for full controllability on
the whole Hilbert space, and often a surprisingly small number of
controls is necessary.  In recent work it was shown that two independent
local controls acting on the first spin suffice for full controllability
for spin chains with coupling of Heisenberg type~\cite{Burgarth-10},
while chains with XX-coupling control of at least the first two qubits
is required \cite{SPP-08,KP-10}.

Alternatively, we would like to know what interesting tasks we can still
perform given certain limited Hamiltonian resources.  These questions
are practically important, because in many real physical systems there
are limitations on the Hamiltonian that can be implemented, and the
quantum dynamics may be restricted to a subspace of the full Hilbert
space~\cite{KP-10,Burgarth,WSBB,Heule-Burgarth-Stojanovic}.  In this
article, we study this subspace controllability problem.  Specifically,
for any spin-$\frac{1}{2}$ network with coupling of XXZ-type subject to
$Z$-directional magnetic control fields, the Hilbert space of the system
always decomposes into so-called excitation subspaces, which remain
invariant under the dynamical evolution~\cite{WSBB}.  Furthermore,
unlike in previous work on restricted local control, which has usually
focused on control of first the spin, we explicitly study the effect of
the position of the controlled node on the controllability on the single
excitation subspace.  We find a surprising relationship between actuator
placement, controllability and dynamical \emph{symmetries} of the system.

Symmetry is an important concept and a powerful tool in
physics~\cite{QM-symmetry}.  Dynamical symmetries provide an alternative
perspective to the controllability problem, implying
non-controllability, and are of interest as characterizing symmetries is
usually easier than calculating the dynamical Lie algebra.  This has
been explored in several works on controllability of spin chains and
networks~\cite{Tannor, Thomas, Peter-Alastair-Sophie, Thomas-2010}.  In
our case we are able to fully characterize all possible dynamical
symmetries for a general XXZ-chain with local $Z$-control of a single
spin in the single excitation subspace with the aid of the Bethe
ansatz~\cite{Bethe}.  While dynamical symmetries imply that the system
is not controllable, lack of symmetry does not always imply
controllability.  This raises the question under which conditions the
converse holds, i.e., lack of symmetry implies
controllability~\cite{Thomas}.  We show that for common models such as
XX and Heisenberg chains and some branched networks lack of symmetry is
not only a necessary but also a sufficient condition for
controllability.  

\section{XXZ Networks with Single-Node Control}
\label{section:model}

\textbf{Intrinsic Hamiltonian:} The Hamiltonian of a network of $N$
spin-$\frac{1}{2}$ particles with XXZ coupling is of the form
\begin{equation}
\label{eqn:H0}
H_s = \frac{1}{2} \sum_{1\le m<n\le N}
      \gamma_{mn} \left( X_{m}X_{n}+Y_{m}Y_{n}+\kappa Z_{m}Z_{n} \right),
\end{equation}
where $X,Y,Z$ are the standard Pauli operators and $X_n$ denotes the
$N$-fold tensor product whose $n$th factor is $X$, all others being the
identity $\ONE$.  The dimensionless constant $\kappa$ determines the
type of coupling such as Heisenberg ($\kappa=1$), XX ($\kappa=0$) or
dipole ($\kappa=-1$) coupling.  The constants $\gamma_{mn}$ determine
the coupling strengths between nodes $m$ and $n$ in the network.
Special cases of interest are chains with \textit{nearest-neighbor}
coupling, for which $\gamma_{mn}=0$ except when $m=n\pm 1$.  A network
is \textit{uniform} if all non-zero couplings are equal, i.e.,
$\gamma_{mn} \in \{0,\gamma\}$.  To every spin network we can associate
a simple graph representation with vertices $\{1,\dots,N\}$ determined
by the spins and edges by non-zeros couplings, i.e., there is an edge
connecting nodes $m$ and $n$ exactly if $\gamma_{mn}\neq 0$.

\begin{figure}
\includegraphics[width=\columnwidth]{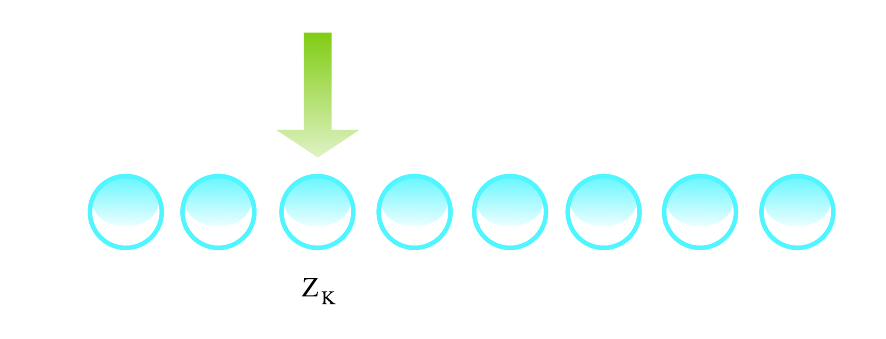}
\caption{XXZ chain with a local Z control $H_1=Z_k$.}
\label{fig:local-Z}
\end{figure}

\textbf{Node Controls:} In this article our main interest are spin
networks subject to a magnetic field in the z-direction or equivalently
a $Z$-control acting \textit{locally}, on a single spin (node/vertex) in
the network as shown in Fig.~\ref{fig:local-Z} for a chain.  The
corresponding control Hamiltonian is
\begin{equation}
   H_c = Z_{k}.
\end{equation}
Neglecting dissipation, the dynamical evolution of the network subject
to a time-varying control field $f(t)$, representing the magnitude of
the magnetic field applied to node $k$, is governed by the Schr\"odinger
equation
\begin{equation}
\label{eqn:dynamics}
  \dot\rho = \tfrac{i}{\hbar}[H_s+f(t)H_c,\rho].
\end{equation}

\textbf{Symmetries:} Following the definitions given
in~\cite{Peter-Alastair-Sophie}, a controlled spin network with
Hamiltonian $H_s+f(t)H_c$ is said to have an \emph{external} or
\emph{commutation symmetry} if there is a Hermitian matrix $S$ which
commutes with both Hamiltonians, i.e., $[H_s,S]=[H_c,S]=0$.  The
existence of such a symmetry operator implies that the Hamiltonians
$H_s$ and $H_c$ can be simultaneously block-diagonalized, and the
Hilbert space decomposed into smaller invariant subspaces corresponding
to the eigenspaces of the symmetry operator $S$.  

The system Hamiltonian of an XXZ spin network always commutes with the
total spin operator $S_F=\sum_n^N (Z_n+I)/2$, $[H_s,S_F]=0$, as does any
control Hamiltonian of the form $Z_k$, and thus any XXZ spin network
with only controls of this type always has this external symmetry.  It
is easy to see that $S_F$ has $N+1$ distinct eigenvalues, ranging from
$n=0$ to $n=N$, corresponding to the possible different numbers of
excitations in the network, and hence we have $N+1$ invariant subspaces.

\begin{figure}
\begin{tabular}{c}
\centerline{\includegraphics{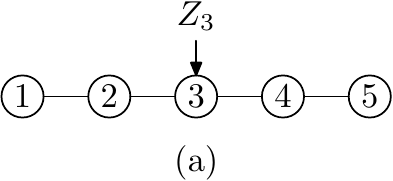}}\\
\centerline{\includegraphics{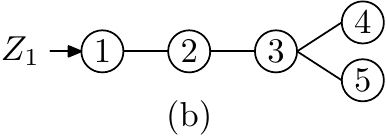}}\\
\centerline{\includegraphics{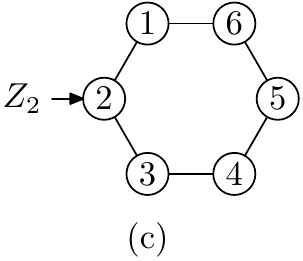}}\\
\centerline{\includegraphics{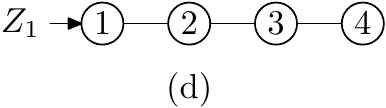}}
\end{tabular}
\caption{Different spin networks with a local Z control at positions
indicated by the arrows. (a),(b),(c) have permutation symmetry, (d) does
not.}  \label{fig1}
\end{figure}

Another common type of external symmetry is \textit{permutation
symmetry}.  A non-identity member of the permutation group $\Pi$ is a
\text{graph symmetry} of the \emph{controlled network} if both $H_s$ and
$H_c$ are invariant under the permutation $\Pi$ of the spins.  For a
single-node control this means in particular that the permutation must
fix the controlled node.  The permutation $\Pi$ induces a permutation
operator $P$ which commutes with both the intrinsic and control
Hamiltonians and hence defines an external symmetry.  In Fig.~\ref{fig1}
there are permutation symmetries in (a), (b) and (c) but not in (d).
Permutation symmetries commute with the total spin operator $S_F$ and
therefore induce symmetries on \emph{all} excitation subspaces.

The existence of external symmetries immediately implies that the system
is not controllable on the full Hilbert space $\H$, i.e., the dynamical
Lie algebra $\LL$ generated by $iH_s$ and $iH_c$ is a proper sub-algebra
of $\su(2^N)$~\footnote{The set of unitary gates that can be implemented
for a dynamical Lie algebra $\LL$ is $e^{\LL}$ and the system is
controllable on the entire Hilbert space if $\LL=\uu(2^N)$ or
$\LL=\su(2^N)$~\cite{D'Alessandro-book}.}.  This is easy to see as the
existence of an external symmetry $S$ means that the system and control
Hamiltonians $H_s$ and $H_c$ are simultaneously blockdiagonalizable and
therefore can not generate $\su(N)$ or $\uu(N)$.  Specifically, if the
Hilbert space decomposes, $\H=\oplus_{d=1}^D \H_d$, where $\H_d$ are
invariant subspaces, then the dynamical Lie algebra generated must be a
subalgebra of $\oplus_{d=1}^D \uu(\dim \H_d)$.  The system may still be
controllable on one or more of the subspace $\H_d$ of $\H$, however, if
the dynamical Lie algebra on the subspace $\H_d$ is equal to $\uu(\dim
\H_d)$ or $\su(\dim \H_d)$.  This is the notion of controllability we
are interested in here, in particular the controllability of the system
on the single excitation subspace $\H_1$ with respect to the total
excitation (symmetry) operator $S_F$.

\section{Subspace Symmetries and Controllability}
\label{section:symmetry}

Denoting the spin-$\frac{1}{2}$ excitation basis vectors by $\ket{\up}$ and
$\ket{\down}$, the single excitation subspace $\H_1$ is spanned by the
$N$ basis vectors: $\ket{\down\up\cdots\up}$, $\ket{\up\down\cdots
\up}$, $\dots$, $\ket{\up\up\cdots\down}$, which we can denote by
$\ket{1},\dots,\ket{N}$, for simplicity.  We can define a basis for the
$N\times N$ antisymmetric matrices as follows:
\begin{align*}
 x_{j,k}&= i(\ket{j}\bra{k}+\ket{k}\bra{j}),\\
 y_{j,k}&= -\ket{j}\bra{k}+\ket{k}\bra{j},\\
 z_{k}  &= i\ket{k}\bra{k},
\end{align*}
with $1\le j<k\le N$.  Restricted to $\H_1$, the Hamiltonians $H_s$ and
$H_c$ are given by the $N\times N$ matrices
\begin{subequations}
 \label{eqn:XXZ1}
 \begin{align}
 H_0 &=-i \sum_{n=1}^{N-1} \gamma_n x_{n,n+1}
       -i \sum_{n=1}^N \mu_n z_n\\
 H_1 &=-i (\ONE - 2 z_{k})
\end{align}
\end{subequations}
where $\gamma_n$ is shorthand for $\gamma_{n,n+1}$ here, and the
diagonal elements are $\mu_n=\mu_0-(\gamma_{n-1}+\gamma_{n})\kappa$
with $\mu_0=\frac{\kappa}{2}\sum_{n=1}^{N-1} \gamma_n$, setting
$\gamma_0=\gamma_{N}=0$ for convenience.

For the special case of a uniformly coupled chain, $\gamma_{n}=1$, we obtain
$\mu_0=(N-1)\tfrac{\kappa}{2}$, $\mu_1=\mu_N=(N-3)\tfrac{\kappa}{2}$ and
$\mu_n=(N-5)\tfrac{\kappa}{2}$ for $1<n<N$, and thus up to a multiple
of the identity we have
\begin{equation}
 \label{eqn:XXZ-H0}
  H_0 =\begin{pmatrix}
         \kappa & 1 & 0 & \cdots & 0 \\
         1 & 0 & 1 & \cdots & 0 \\
         \cdots  & \cdots  & \cdots  & \cdots  & \cdots \\
         0 & \cdots &  1 & 0 & 1\\
         0 & \cdots  & 0 & 1 & \kappa
        \end{pmatrix},\\
\end{equation}
$H_1$ is a diagonal matrix with ones everywhere except for the $k$
diagonal element, which is $-1$.  Subtracting again and changing
sign we can equivalently take
\begin{equation}
 \label{eqn:XXZ-H1}
 H_1 =\begin{pmatrix}
        0 & \cdots & 0 & \cdots & 0  \\
       \cdots  & \cdots  & \cdots  & \cdots  & \cdots \\
       0 & \cdots & 1 & \cdots &  0\\
       \cdots  & \cdots  & \cdots  & \cdots  & \cdots \\
       0 & \cdots & 0 & \cdots & 0
      \end{pmatrix}.
\end{equation}
Note that the addition of multiples of identity matrix to $H_s$ and
$H_c$ does not change their commutation relations with the original
Hamiltonian, and hence the Lie algebra generated by $H_0$ and $H_1$
differs from the Lie algebra generated by the original Hamiltonians at
most by the identity $\ONE$.

In the following, let $\LL$ denote the dynamical Lie algebra generated
by $H_0$ and $H_1$.  If $\dim(\LL)=N^2$ or $N^2-1$ then $\LL=\uu(N)$ or
$\su(N)$ and the system is controllable on $\H_1$.  The generators
$x_{k,j},y_{k,j},z_{k}$ defined above form a natural basis for the Lie
algebra $\u(N)$, and if we can generate them from $-iH_0=\sum_j^{N-1}
x_{j,j+1}$ and $-iH_1=z_k$, then $\LL=\u(N)$ and the system is
controllable.  We shall also use the following commutation relations:
\begin{align*}
[x_{jk},z_k]      &= y_{jk}\\
[y_{jk},z_k]      &=-x_{jk}\\
[x_{jk},x_{k\ell}]&= y_{j\ell}\\
[x_{jk},y_{k\ell}]&=-x_{j\ell}\\
[x_{jk},y_{jk}]   &= 2(z_j-z_k)
\end{align*}
Moreover, to show controllability it suffices to show
that we can generate all $x_{n,n+1}$ and $y_{n,n+1}$ for $n=1,\ldots,N-1$.


As noted above, permutation symmetries commute with the total spin
operator $S_F$ and therefore induce symmetries on all eigenspaces of $S_F$,
including $\H_1$.  Thus, for the XXZ
networks shown in Fig.~\ref{fig1} we can immediately conclude that the
system is not controllable on any excitation subspace (except the
trivial ones $n=0$ and $n=N$) if (a) we control the middle spin, (b) we
control any spin outside the two equal-length branches, and (c) we
control any spin on the ring.  However, permutation symmetries are not
the only possible symmetries, especially on $\H_1$.

\textbf{Example:}  For an XX chain ($\kappa=0$) of length $N=7$ and $k=2$, we
find that both $H_0$ and $H_1$, restricted to the single excitation
subspace, commute with
\begin{equation*}
M = \begin{pmatrix} 0&0&1&0&-1&0&1\\
                0&1&0&0&0&0&0\\
                1&0&0&0&1&0&-1\\
                0&0&0&1&0&0&0\\
                -1&0&1&0&0&0&1\\
                0&0&0&0&0&1&0\\
                1&0&-1&0&1&0&0
     \end{pmatrix},
\end{equation*}
which is not a permutation symmetry.  We can indeed verify that the
dimension of the Lie algebra is $\dim(\LL)=36<7^2$, i.e., the system is
not controllable on $\H_1$, and the restrictions of $H_0$ and $H_1$ to
$\H_1$ can be simultaneously block-diagonalized.

Firstly, we can now show generally that for systems with a single-node
$Z$-type control, the external symmetries can be very easily characterized.

\begin{theorem}
\label{thm:vk0}
An XXZ spin network with a single node control $Z_k$ has external
symmetries on $\H_1$ if and only if $H_0$ has one or more eigenvectors
$\ket{v}$ with zero overlap with node $\ket{k}$, i.e., $\ip{k}{v}=0$.
\end{theorem}

\begin{IEEEproof}
Without loss of generality, we can choose a basis such that $H_0$ is
diagonal.  From Favard's theorem~\cite{Chihara} for tridiagonal
matrices, we know that the eigenvalues of $H_0$ are distinct.  In such a
basis $H_1$ takes the form $B_{k\ell}=\ip{v_k}{1}\ip{1}{v_\ell}$ and
setting $C=[M,B]$ gives $C_{k\ell}=(m_k-m_\ell)B_{k\ell}$, which shows
that $C_{k\ell}$ can vanish only if $B_{k\ell}=0$ or $m_k=m_\ell$.  If
all $B_{k\ell}\neq 0$ then $C_{k\ell}=0$ for all $k,\ell$ is only
possible if $m_k=m_\ell$ for all $k,\ell$, i.e., if $M$ is a multiple of
the identity, in which case there is no symmetry.  Thus a symmetry exists
if and only if $B_{k\ell}=0$ for some $k,\ell$, which is equivalent to
$\ip{v_k}{1}=0$ for some $k$, i.e., the existence of an eigenvector
$\vec{v_k}$ that has zero overlap with the controlled node.
\end{IEEEproof}

\cite{Burgarth-07} found a similar characterization in the setting of 
control by relaxation, and \cite{Peter-Alastair-Sophie} showed the
existence of external symmetries to be equivalent to the existence of
eigenstates of the system Hamiltonian that have no overlap with the
pendant vertex for spin networks with a pendant-type control.

We shall refer to systems with no external or commutation symmetries as
\emph{indecomposable}.  Note that spin networks given by coupling graphs
with multiple disjoint components are always decomposable as are systems
with permutation symmetries, but as we have seen, these are by no means
the only external symmetries the system might have.  External symmetries
imply that the Hilbert space can be written as a direct sum of subspaces
that are invariant under the dynamics.  Assuming we have found all
external symmetries and decomposed the Hilbert space into invariant
subspaces that are not further reducible, the dynamics on each invariant
subspaces can be subject to \emph{internal Lie group symmetries},
characterized by the existence of a \emph{unitary} or
\emph{anti-unitary} operator $S$ such that
\begin{equation}
   (iH_0)^T S + S (i H_0) = (i H_1)^T S + S (i H_1) = 0,
\end{equation}
where $H_0$ and $H_1$ are required to be trace-zero and $A^T$ denotes
the transpose of $A$.  These internal symmetries can be divided into
orthogonal and symplectic symmetries, and the existence of such a
symmetry implies that the dynamical Lie algebra $\LL$ generated by $H_0$
and $H_1$ is a subalgebra of $\so(N)$ or $\sp(N)$, respectively, and
thus the system is again not controllable on the respective subspace.
For spin networks with a particular single controlled coupling (a
\textit {pendant control}) it was recently found that internal
symmetries on the single excitation subspace $\H_1$ are of orthogonal
type and related to the existence of a bipartite structure of the
network~\cite{Peter-Alastair-Sophie}.  In contrast, internal symmetries
never occur for XXZ spin networks with single-node $Z$-controls.

\begin{theorem}
An indecomposable XXZ network with a single-node Z-control $-iH_1=z_k$
does not permit internal symmetries on $\H_1$.
\end{theorem}

\begin{IEEEproof}
Let $\bar{H}_m=H_m-\frac{1}{N}\Tr(H_m)H_m\ONE_N$ for $m=0,1$, be the
zero-trace versions of the Hamiltonians $H_0$ and $H_1$.  We need to
show there is no internal symmetry between $\bar H_0$ and $\bar H_1$.
As our Hamiltonians are both real-symmetric the internal symmetry
condition can be simplified
\begin{equation*}
  \bar{H}_0 S + S \bar{H}_0=0, \quad \bar{H}_1 S + S \bar{H}_1=0.
\end{equation*}
We see that when $\bar{H}_1$ is diagonal, $\bar{H}_1=\diag(a_n)$, the
latter condition is equivalent to $(a_{m}+a_n)S_{mn}=0$, which implies
$S_{mn}=0$ unless $a_m=-a_n$.  In particular, if $H_1=z_k$ then we have
$a_k=1$ and $a_n=-\tfrac{1}{N-1}$ for $n\neq k$ and thus the sum of any
two diagonal elements of $\bar{H}_1$ never vanishes.
\end{IEEEproof}

Thus, remarkably for this type of control, we only need to consider
external symmetries, which we shall refer to simply as symmetries in the
following.  In fact, the proof of the previous theorem shows that a
decomposable spin network with $Z$-controls admits internal symmetries
on the single excitation subspace if only if $N$ is even and we
collectively control exactly half of all nodes, in which case
$\bar{H}_1$ will contain equal numbers of $+1$ and $-1$ entries, which
can cancel.

\begin{theorem}
\label{thm:z1_control} An XXZ chain with arbitrary coupling strengths
with a single Z control at the end node is controllable on the
single-excitation subspace $\H_1$.
\end{theorem}

\begin{IEEEproof}
To show controllability we calculate the dynamical Lie algebra $\LL$
generated by $-iH_0$ and $-iH_1$, noting that the Hamiltonians take the
form~(\ref{eqn:XXZ1}) and (\ref{eqn:XXZ-H1}) with $\gamma_n\ne 0$ for
any $1\le n<N$ and $k=1$.  The commutators
\begin{align*}
 \gamma_1^{-1}[-iH_0,z_1]            &= y_{1,2}\\
 [z_1,y_{1,2}]                       &= x_{1,2}, \\
 z_1 + \tfrac{1}{2}[x_{1,2},y_{1,2}] &= z_2
\end{align*}
immediately give us the three generators of the Lie algebra.  Using
these generators we can define a reduced system generated by $-iH_2$ and
$z_2$ with
\begin{equation*}
 -iH_2 = -iH_0-x_{1,2}-\mu_1z_1
     = \sum_{k=2}^{N-1} \gamma_k x_{k,k+1}+\sum_{k=2}^{N}\mu_k z_k,
\end{equation*}
which represents an XXZ chain of length $N-1$.  By the same procedure as
above, we can now generate $x_{2,3}$, $y_{2,3}$ and $z_3$; iterating the
procedure $N-2$ times, we can generate $x_{n,n+1}$, $y_{n,n+1}$ and
$z_n$ for $n=1,\ldots,N-1$, i.e., all the generators corresponding to
the simple roots of the Lie algebra $\uu(N)$.
\end{IEEEproof}

Controllability implies the non-existence of symmetries but we can see
directly that there are no external symmetries.  Theorem~\ref{thm:vk0}
shows that the first entry of any eigenvector of $H_0$ cannot be zero.
Suppose we have a eigenvector $\vec{v}$ of $H_0$ with $v_1=0$.  Then
$H_0\vec{v}=\lambda \vec{v}$ gives
\begin{align*}
  \begin{pmatrix} \mu_1 & \gamma_1 & 0 & \cdots & 0 \\ \gamma_1 &
\mu_2 & \gamma_2 &
\cdots & 0 \\ \vdots & \ddots & \ddots & \ddots & \vdots \\ 0 & \cdots &
\gamma_{N-2} & \mu_{N-1} & \gamma_{N-1}\\ 0 & \cdots & 0 & \gamma_{N-1} &
\mu_{N} \end{pmatrix}
\begin{pmatrix} 0 \\ v_2 \\ \vdots \\ v_{N-1}\\ v_{N}
 \end{pmatrix}
 =\begin{pmatrix} \gamma_1v_2 \\ \gamma_2v_3 \\ \vdots \\ \vdots \end{pmatrix},
\end{align*}
which in turn implies $v_2=v_3=\cdots=0$, i.e. $\vec v=0$.  Thus any
eigenvector of $H_0$ cannot be zero at its first entry, i.e., there is
no external symmetry.

This end-controllability result can be generalized to the case where we
collectively control $k$ spins from the end of the chain, i.e.~$-iH_1 =
\sum_{j=1}^{k<N}z_j$.

\begin{theorem}
For XXZ chains of length $N$ with Z-control of $k<N$ spins at one end of
the chain, the system is controllable and hence has no symmetry on $\H_1$.
\end{theorem}

\begin{IEEEproof}
Given Theorem \ref{thm:z1_control}, we only need to show that from $-iH_0$
and $-iH_1=z_1+\cdots+z_{k}$, $k<N$, we can generate $z_1$, and then the
controllability follows.  We have
\begin{align*}
[-iH_1,-iH_0]         &\rightarrow y_{k,k+1}\\
[y_{k,k+1},-iH_1]     &\rightarrow x_{k,k+1}\\
[x_{k,k+1},y_{k,k+1}] &\rightarrow z_{k}-z_{k+1}
\end{align*}
and defining
\begin{equation*}
-iH_0' \equiv -iH_0- \gamma_k x_{k,k+1}
                     + \tfrac{1}{2}(\mu_k-\mu_{k+1}) (z_k-z_{k-1}) \\
\end{equation*}
we further have
\begin{align*}
[-iH_0',y_{k,k+1}]    &\rightarrow -x_{k-1,k+1}+x_{k,k+2}\\
[-iH_0',x_{k,k+1}]    &\rightarrow -y_{k-1,k+1}+y_{k,k+2}\\
[-x_{k-1,k+1}+x_{k,k+2},&-y_{k-1,k+1}+y_{k,k+2}]\\
 &\rightarrow (z_{k+2}-z_{k+1})-(z_{k}-z_{k-1})
\end{align*}
as well as
\begin{align*}
[-iH_0',z_{k+1}-z_{k}]                &\rightarrow y_{k+1,k+2}+y_{k-1,k}\\
[z_{k+1}-z_{k},y_{k+1,k+2}+y_{k-1,k}] &\rightarrow x_{k+1,k+2}+x_{k-1,k}\\
[x_{k+1,k+2}+ x_{k-1,k}, y_{k+1,k+2}+ y_{k-1,k}]&\\
 \rightarrow (z_{k+2}- &z_{k+1})+(z_{k}-z_{k-1})
\end{align*}
Thus, we can generate $z_{k+2}-z_{k+1}$ and $z_{k}-z_{k-1}$.

Moreover, from $[z_{k}-z_{k-1},-iH_0']$ we can get $y_{k-1,k}$ and we
get $x_{k-1,k}$. Hence, we find the following rule: starting from
$z_{k+1}-z_{k}$, $x_{k,k+1}$ and $y_{k,k+1}$, we can generate
$z_{k+2}-z_{k+1}$ and $z_{k}-z_{k-1}$. Then, starting from
$z_{k}-z_{k-1}$, $x_{k-1,k}$, and $y_{k-1,k}$, we can analogously
generate $z_{k-1}-z_{k-2}$, $x_{k-2,k-1}$, and $y_{k-2,k-1}$. Repeating
this process, we can sequentially generate $z_{k}-z_{k-1}$,
$z_{k-1}-z_{k-2}$, $\ldots$, $z_2-z_1$, and together with $-iH_1$, we
can finally generate $z_1$, and the controllability follows.
\end{IEEEproof}

Based on these results one might conjecture that we always have subspace
controllability, at least for inhomogeneous chains, regardless of which node we control.  But this is not
the case, and in fact some symmetries are extremely robust even in the
presence of inhomogeneity.

\begin{theorem}
An XX chain of odd length always has an external symmetry (and is thus
not controllable) on $\H_1$ if the index $k$ of the controlled spin is
\emph{even}.
\end{theorem}

\begin{IEEEproof}
The Hamiltonian of XX chain reduces to
\begin{equation*}
  -iH_0 = \sum_{n=1}^{N-1} \gamma_k x_{n,n+1}
\end{equation*}
Favard's theorem~\cite{Chihara} for tridiagonal matrices guarantees that
$H_0$ has $N$ distinct eigenvalues $\gamma_j$ with corresponding
eigenvectors $\vec{v}_j=(v_{kj})$ satisfying
\begin{equation*}
  v_{kj} = (-1)^{k-1}\frac{f_{k-1}(\lambda_j)}{\gamma_1\cdots\gamma_{k-1}},
\end{equation*}
where the polynomial $f_{n+1}(\lambda)$ is determined by the recurrence
relation:
\begin{equation*}
  f_{n+1}(\lambda)=\lambda f_{n}(\lambda)-\gamma_n^2f_{n-1}(\lambda),
\end{equation*}
for $n>0$ and $f_0=1$ and $f_{-1}=0$.  Moreover, $f_{N}(\lambda)$ is the
characteristic polynomial of $H_0$.  If $N$ is odd then $f_N$ is an odd function too, so $\lambda=0$ is a root of $f_N(\lambda)$ with eigenvector $\vec{v}_0
=(f_0(0),f_1(0),\dots)$ and $f_1(0)=f_3(0)=\dots=f_{N}(0)=0$, i.e. all
even entries vanish. Hence if $k$ is even, there is an external symmetry
and the system is not controllable.
\end{IEEEproof}

A general XXZ chain, either homogeous or not, is called centro-symmetric
with respect to the centre of the chain if the couplings $\gamma_{n}
:=\gamma_{n,n+1}$ between adjacent spins satisfy $\gamma_{n}=
\gamma_{N-1-n}$ for $n=1,\ldots, N-1$.

\begin{theorem}
A centro-symmetric XXZ chain is not controllable on the $\H_1$ subspace
if $N=2k-1$, where $k$ is the controlled spin.
\end{theorem}

\begin{IEEEproof}
For a centro-symmetric chain of length $N=2k-1$ the characteristic
polynomial is of the form $f_N(\lambda)=f_{k-1}(\lambda) p(\lambda)$,
where both $f_{k-1}(\lambda)$ and $p(\lambda)$ are polynomials of degree
$k-1$ and $k$, respectively.  This shows that $k-1$ of the roots
$\lambda_m$ of $f_N(\lambda)$ must be roots of $f_{k-1}(\lambda)$, and
hence $f_{k-1}(\lambda_m)=0$ for these $\lambda_m$, showing that the
corresponding eigenvectors are $\ket{\lambda_m}$ are ``dark states,''
i.e., satisfy $[H_S+f(t)H_c]\ket{\lambda_m}=0 \ket{\lambda_m}$ for any
control $f(t)$, if we control the middle spin $k=(N+1)/2$.  Thus the
maximum controllable subspace is has dimension $k+1$.
\end{IEEEproof}


\section{Characterization of Symmetries for Homogeneous XXZ Chains}


From Theorem~\ref{thm:vk0}, $H_0$ and $H_1$ have an external symmetry if
and only if there exists some eigenvector $\vec{v}_j=(v_{kj})$ of $H_0$
such that $v_{kj}=0$.  For a homogeneous chain the Hamiltonian $H_0$ is
tridiagonal with uniform values for the off-diagonal elements and zeros
on the diagonal except for the first and last entry, and if we define
\begin{equation}
\label{eqn:a0_a1}
 v_0     = \kappa v_1, \quad
 v_{N+1} = \kappa v_{N}.
\end{equation}
then the eigenvalue equation $H_0\vec{v}=E\vec{v}$ can be written as
$v_{k-1}+v_{k+1}=Ev_{k}$ for $k=1,\dots,N$.  This suggests that the
eigenvectors $\vec{v}$ are of the form (Bethe ansatz~\cite{Karabach})
\begin{equation}
  \label{eqn:v}
  v_k = Az^k + Bz^{-k}
\end{equation}
with $z=e^{i\theta}$.  Substituting into (\ref{eqn:a0_a1}) gives:
\begin{subequations}
 \label{eqn:A}
\begin{align}
  A+B                  &= \kappa (Az+Bz^{-1})\\
  Az^{N+1}+Bz^{-(N+1)} &= \kappa (Az^{N}+Bz^{-N}).
\end{align}
\end{subequations}
The first equation gives $B=A z \tfrac{1-\kappa z}{\kappa -z}$,
assuming $A\neq 0$, $z\neq 1$.  Inserting this into the second
equation we obtain
\begin{equation}
\label{eqn:z2N}
 z^{2N} = \frac{(1-\kappa z)^2}{(\kappa -z)^2} \quad\Rightarrow\quad
 z^{N}  = \pm \frac{1-\kappa z}{\kappa -z}.
\end{equation}
The condition for the existence of a symmetry, $v_k=0$, thus becomes
$A z^{2k}+B =0$, and inserting $B=A z \tfrac{1-\kappa z}{\kappa -z}$,
\begin{equation}
  \label{eqn:z_2k}
  z^{2k-1}=-\frac{1-\kappa z}{\kappa- z}=\pm z^N.
\end{equation}
Equation~(\ref{eqn:z_2k}) allows us to find all possible values of
$\kappa$ that allow external symmetries for given $N$ and $k$.  If
$N=2k-1$, for instance, then any $\kappa$ satisfies (\ref{eqn:z_2k}) and
the system always has a symmetry.  Indeed, this is the case of an odd
chain where we control the central spin, and it is easy to see that this
system has a permutation symmetry as discussed before.  Similarly, if
$N=2k$ then there are no symmetries for any $\kappa$ as $z^N=\pm
z^{N-1}$ can be satisfied only if $z=0,\pm 1$ but these solutions do not
correspond to valid Bethe eigenvectors.

More generally, substituting $z=e^{i\theta}$ into (\ref{eqn:z_2k}) gives
\begin{equation}
 \label{eq:sol}
  e^{i\phi}\equiv e^{i(2k-1)\theta}
  = -\frac{1-\kappa z}{\kappa-z}
  =  \frac{(\kappa \cos\theta-1)+i\kappa\sin\theta}
          {(\kappa-\cos\theta)- i\sin\theta}.
\end{equation}
This equation contains two independent equations for the real and
the imaginary parts:
\begin{align*}
\cos\phi (\kappa-\cos\theta) + \sin\phi\sin\theta
 &= \kappa\cos\theta-1 \\
\sin\phi (\kappa-\cos\theta) - \sin\theta\cos\phi
 &= \kappa\sin\theta.
\end{align*}
The solutions of these two equations are, respectively,
\begin{align*}
  &\kappa=\frac{\cos(\theta+\phi)-1}{\cos\phi-\cos\theta},
  & \cos\phi\ne\cos\theta \\
  &\kappa=\frac{\sin(\phi+\theta)}{\sin\phi-\sin\theta},
  &\sin\phi\neq \sin\theta.
\end{align*}
Substituting $\phi=(2k-1)\theta$ and using elementary trigonometric
identities, both of these solutions simplify to
\begin{equation}
\label{eqn:kappa}
  \kappa =\frac{\sin(k\theta)}{\sin(k-1)\theta}.
\end{equation}
Substituting $z=e^{i\theta}$ into Eq.~(\ref{eqn:z_2k}) shows further
that we must have $e^{i(2k-1)\theta}=\pm e^{iN\theta}$, or
$e^{i(N-2k+1)\theta}=\pm 1$.  For $2k-1<N$ this gives
\begin{equation}
  \label{eqn:theta}
  \theta = \frac{j\pi}{N-(2k-1)}, \quad j\in \ZZ.
\end{equation}
Thus, for given $N$ and $k$, from (\ref{eqn:theta}) we can find the
corresponding $\theta$ and $\kappa$ such that the system has a symmetry.
We can traverse all possible values of $N$, $k$ and $j$ in order to find
all types of homogeneous XXZ chain that has an external symmetry. A set
of these values are summarized in Table~\ref{tab:sym} for small values
of $N$. An immediate consequence of this is:

\begin{theorem}
There is at most a countably-infinite number of $\kappa$ that permit
external symmetry. 
\end{theorem}

The theorem implies that for a generic $\kappa$ uniformly chosen at
random from the real line, there will be no symmetry for any $N$ and
$k$.  We must be careful, however, with results such as this because the
most common values for $\kappa$ for real physical systems are $\kappa=0$
(XX-coupling), $\kappa=1$ (Heisenberg), and $\kappa=-1$ (dipole
coupling) and for these special values of $\kappa$ symmetries exist for
many choices of $N$ and $k$.

For $\kappa=0$ (\textbf{XX-coupling}) Eq.~(\ref{eqn:z2N}) gives
$z^{N+1}=\pm 1$ with $z=e^{i\theta}$ or $\theta=\tfrac{j\pi}{N+1}$,
Eq.~(\ref{eqn:A}a) gives $A=-B$ and (\ref{eqn:v}) thus becomes
\begin{equation}
\label{eqn:XX_eigenvector}
  v_{kj} = A e^{i k\theta_j} - A e^{-ik\theta_j}
         = C \sin(k\theta_j)
\end{equation}
for $j,k=1,\ldots,N$ with corresponding eigenvalues are $E_j =
2\cos(\theta_j)$. A simple calculation reveals the normalization
constant to be $C=\sqrt{2/(N+1)}$.  For the system to have an external
symmetry the $k$th entry of one of the eigenvectors $\vec{v}_{j}$ must
vanish.  In this case this happens only if $\sin(k\theta_j)=0$, or
$k\theta_j= \ell\pi$, i.e., if there exists an integer $\ell>0$ such
that $kj=(N+1)\ell$, or equivalently if and only if $N+1$ and $k$ have
a common divisor $>1$ or $\gcd(N+1,k)=g>1$, where $\gcd(\cdot,\cdot)$
represents the greatest common divisor.  Hence, the Hamiltonians have an
external symmetry if and only if $\gcd(N+1,k)=g>1$.

Similarly, for $\kappa=1$ \textbf{(Heisenberg coupling)}
Eq.~(\ref{eqn:z2N}) gives $z^{2N}=1$ and we
obtain~\cite{Bose-Heisenberg}:
\begin{equation}
  v_{kj}= \cos((2k-1) \theta_j), \quad \theta_j = \frac{j\pi}{2N},
\end{equation}
which shows that the system has an external symmetry if and only if
$\gcd(N,2k-1)=g>1$.  This is identical to the result for $\kappa=-1$
\textbf{(dipole-coupling)}\footnote{This result applies to a linear
chain with nearest-neighbor coupling, which is somewhat artificial for a
dipole-coupled chain as dipole coupling tends to be long-range.}.  This
is true is general, owing to the fact there is a 1-to-1 correspondence
between the eigenvalues and eigenvectors of XXZ chains with opposite
$\kappa$.

\begin{theorem}
Homogeneous XXZ chains with opposite values of $\kappa$ have the same external
symmetries.
\end{theorem}

\begin{IEEEproof}
Let the Hamiltonian of an XXZ chain be $H_0=H_0[\kappa]$.  Assuming it
satisfies the eigenvalue equation: $H_0\vec{v}=E\vec{v}$, and defining
$P=\Pi_{j=1}^{N/2} Z_{2j}$, we have $(PH_0P^\dag) P\vec{v}=EP\vec{v}$, and
$PH_0P^\dag =-(XX+YY)+\kappa ZZ=-H_0[-\kappa]$.  Hence, if $E$ and
$\vec{v}$ are the eigenvalue and the eigenvector of $H_0[\kappa]$, then
$-E$ and $P\vec{v}$ are the eigenvalue and the eigenvector of
$H_0[-\kappa]$.  Moreover, as $P\vec{v}=[v_1,-v_2,v_3,-v_4,\cdots,]^T$,
if $N$ and $k$ are chosen such that $v_k=0$ then the kth entry of
$P\vec{v}$ is also equal to zero.  Hence, $H_0[\kappa]$ and
$H_0[-\kappa]$ have the same external symmetries for given $N$ and $k$.
\end{IEEEproof}

\begin{table}
\begin{tabular}{llll}
$N$& $k$ &  $\theta=\frac{j\pi}{N-(2k-1)}$ & $\kappa$\\
\hline
5 & 2 & $\frac{\pi}{2}$ & 0\\
6 & 2 & $\frac{\pi}{3},\frac{2\pi}{3}$ & $\pm 1$\\
6 & 3 &                 & No solution\\
7 & 2 & $\frac{\pi}{2}$ & 0\\
7 & 3 &                 & No solution\\
8 & 2 & $\frac{\pi}{5},\frac{2\pi}{5}$ & $\pm 2\cos\frac{\pi}{5}$\\
8 & 3 & $\frac{\pi}{3}$ & 0\\
8 & 4 &                 & No solution\\
9 & 2 &
$\frac{\pi}{6},\frac{5\pi}{6},\frac{\pi}{3},\frac{2\pi}{3},\frac{\pi}{2}$ &
$\pm \sqrt{3},\pm 1,0$\\
9 & 3 & $\frac{\pi}{4},\frac{3\pi}{4}$ & $\pm \frac{\sqrt{2}}{2}$\\
9 & 4 & $\frac{\pi}{2}$ & 0\\
10 & 2 & $\frac{j\pi}{7},j=1,\dots,6$ & $\pm 2\cos\frac{\pi}{5},\pm
2\cos\frac{2\pi}{5},\pm 2\cos\frac{3\pi}{5}$\\
10 & 3 & $\frac{\pi}{5},\frac{2\pi}{5}$ & $\pm 1$\\
10 & 4 &  & No solution\\
10 & 5 &  & No solution\\
\end{tabular}
\caption{For different values of $N$ and $k$,
$\theta$ and $\kappa$ which make $v_{jk}=0$ can be calculated. }
\label{tab:sym}
\end{table}

\section{Characterization of $\H_1$-Controllability for uniform XX and 
Heisenberg Chains}

In the previous section we characterized the symmetries for XXZ chains,
and we know that the absence of symmetry is a necessary condition for
controllability.  Unfortunately, it is not sufficient in general.

\begin{figure}
\center\includegraphics{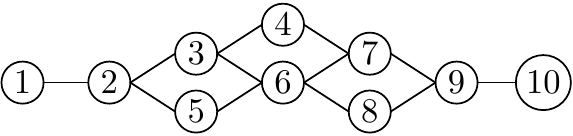} 
\caption{An XX spin network with $N=10$ spins. On $\H_1$, we consider
multi-node Z control, with (a) $H_1=z_1$, (b) $H_1=z_1+z_2$, (c)
$H_1=z_1+z_2+z_3$ and (d) $H_1=z_1+z_2+z_3+z_4$.}
\label{fig:2nd-excitation}
\end{figure}

\textbf{Example.} Consider an XX-spin network composed of $N=10$ spins
as illustrated in FIG.~\ref{fig:2nd-excitation} with a $Z$-control
applied (jointly) to several spin nodes: (a) $H_1=z_1$, (b)
$H_1=z_1+z_2$, (c) $H_1=z_1+z_2+z_3$ and (d) $H_1=z_1+z_2+z_3+z_4$.
Calculating the dynamical Lie algebra $\LL$ generated by $iH_0$ and
$iH_1$, we have in (a) and (b) $\dim(\LL)=81$; in (c) $\dim(\LL)=100$;
in (d) $\dim(\LL)=25$, i.e., only in case (c) do we have $\H_1$-subspace
controllability.  The result in the first two cases is due to the
existence of an external symmetry, in this case a single dark state,
i.e., an eigenstate $\vec{v}$ of $H_0$ that has no overlap with the
controlled spin, $\ip{k}{v}=0$, and a controllable subspace of dimension
$9$.  In case (d), however, one can verify that no symmetries exist, and
the Lie algebra generated is an irreducible representation of $\uu(5)$.
This result can be explained if we realize that the first excitation
subspace Hamiltonians $H_0$ and $H_1$ for this network are in fact
identical to the second excitation subspace Hamiltonians for a uniform
linear chain of length $N=5$, and the Lie algebra for this system is
indeed $\uu(5)$ with the second excitation subspace corresponding to the
10-dimensional anti-symmetric (irreducible) representation of $\uu(5)$.

\textbf{Example.}  The above correspondence between the graph in 
FIG.~\ref{fig:2nd-excitation} and the second excitation subspace
Hamiltonian of an XX chain of length $N=5$ also holds for the 
inhomogeneous case.  For example, the single excitation subspace
Hamiltonians for a $10$-spin network
\begin{align*}
 H_0 = \begin{pmatrix}
    0 & 2 & 0 & 0 & 0 & 0 & 0 & 0 & 0 & 0\\
    2 & 0 & 3 & 0 & 1 & 0 & 0 & 0 & 0 & 0\\
    0 & 3 & 0 & 4 & 0 & 1 & 0 & 0 & 0 & 0\\
    0 & 0 & 4 & 0 & 0 & 0 & 1 & 0 & 0 & 0\\
    0 & 1 & 0 & 0 & 0 & 3 & 0 & 0 & 0 & 0\\
    0 & 0 & 1 & 0 & 3 & 0 & 4 & 2 & 0 & 0\\
    0 & 0 & 0 & 1 & 0 & 4 & 0 & 0 & 2 & 0\\
    0 & 0 & 0 & 0 & 0 & 2 & 0 & 0 & 4 & 0\\
    0 & 0 & 0 & 0 & 0 & 0 & 2 & 4 & 0 & 3\\
    0 & 0 & 0 & 0 & 0 & 0 & 0 & 0 & 3 & 0
\end{pmatrix}
\end{align*}
with collective $Z$-control of the first four spins, $H_1=z_1+z_2+z_3+z_4$,
is equivalent to the second excitation subspace Hamiltonian of a chain of 
length $N=5$ where the couplings between adjacent spins are
$\gamma_{1}:\gamma_{2}:\gamma_{3}:\gamma_{4}=1:2:3:4$ and we control the
first spin only, and we find $\dim(\LL)=25<10^2$, or $\LL=\uu(5)$ and no
symmetries either. This shows that even networks with non-uniform coupling
without any symmetry can be non-controllable.

From the above examples, we see that there are XXZ spin networks without
any symmetries in $\H_1$ that are nonetheless not controllable on this
subspace.  Nevertheless, we shall show that for a particular type of
spin network, the linear chain with a single controlled node, lack of
symmetry is not only necessary but also sufficient for controllability
on $\H_1$.  We shall give rigorous proofs for XX and Heisenberg chains,
which are of most practical interest, but the same techniques could be
applied to prove controllability for other types of chains.

\subsection{XX Chain}

\begin{theorem}
\label{thm:XX} For an XX chain of length $N$ with local control $z_k$,
the system is controllable on $\H_1$ if and only if $\gcd(N+1,k)=1$.
\end{theorem}

Note that this condition is equivalent to the no-symmetry condition
derived in the previous section.

\begin{IEEEproof}
We already know that that the system is controllable on the single
excitation subspace for $k=1$.  Also, by symmetry, we only need to
discuss the case when $1\le k\le N/2$.  Assuming $1<k\le N/2$, we would
like to determine all the operators we can generate from $H_0$ and $H_1$
through commutation relations.

\begin{result}
\label{result:nk} For an XX chain on $\H_1$ with $H_1=-iz_k$ and
$N=rk+m$, $0\le m<k$, we can generate $z_{qk}$ with $1\le q\le r$,
starting from $z_k$.
\end{result}

We can generate the following elements:
\begin{align*}
[-iH_0,z_k]&\rightarrow y_{k-1,k}+y_{k+1,k}\\
[y_{k-1,k}+y_{k+1,k},z_k]&\rightarrow x_{k-1,k}+x_{k+1,k}\\
-iH_2 &\equiv-iH_0-(x_{k-1,k}+x_{k+1,k})\\
[-iH_2,x_{k-1,k}+x_{k+1,k}]&\rightarrow y_{k-2,k}+y_{k+2,k}\\
[y_{k-2,k}+y_{k+2,k},z_k]&\rightarrow x_{k-2,k}+x_{k+2,k}\\
\cdots\cdots &\cdots \cdots\cdots\\
[-iH_2,x_{2,k}+x_{2k-2,k}]&\rightarrow y_{1,k}+y_{2k-1,k}\\
[y_{1,k}+y_{2k-1,k},z_k]&\rightarrow x_{1,k}+x_{2k-1,k}\\
[-iH_2,x_{1,k}+x_{2k-1,k}]&\rightarrow y_{2k,k}\\
[y_{2k,k},z_k]&\rightarrow x_{2k,k}\\
[y_{2k,k},x_{2k,k}]&\rightarrow z_{2k}
\end{align*}
Thus, starting from $-iH_0$ and $z_k$, we can generate $z_{2k}$, and
continuing this process, we can sequentially generate
$z_{3k},z_{4k},\ldots$. 

For $N=rk$ this implies that we can generate $z_N$, which is equivalent
to $z_1$ and implies controllability by Theorem~\ref{thm:z1_control}.

\begin{result}
\label{result:diff} If we can generate $z_{k_1}$ and $z_{k_2}$ with
$k_1-k_2=1$, staring with $z_k$ and $H_0$, then the system is
controllable.
\end{result}

If we can generate two local operators $z_{k_1}$ and $z_{k_2}$ with
$k_1-k_2=1$ from $z_k$ and $H_0$ then we can further generate
\begin{align*}
  H_0' = x_{1,2}+\cdots+x_{k_1-2,k_1-1}+x_{k_1+1,k_1+2}+\cdots+x_{N-1,N},
\end{align*}
and together with $z_{k_1-1}$ we can sequentially generate $z_{k_1-2}$,
$z_{k_1-3},\ldots$ until we obtain $z_1$, analogous to the proof of
Theorem~\ref{thm:z1_control}, and Theorem~\ref{thm:z1_control} itself
then again implies controllability.

\begin{result}
\label{result:k-m} Let $N=rk+m$, $0\le m<k$. If $r=2r'$ is even, then
$z_k$ and $H_0$ can generate $z_m$; If $r=2r'+1$ is odd, then $z_k$ and
$H_0$ can generate $z_{k-m}$.
\end{result}

With the above results we can apply the Euclidean algorithm to complete
the proof of Theorem~\ref{thm:XX}.

Let $n_1=k$. For $\gcd(N+1,n_1)=1$, and $n_1\mid(N+1)$, we have
$N+1=r_1n_1+m_2$, $0<m_2<n_1$.  By Result~\ref{result:k-m}, starting
with $z_{n_1}$, we can generate $z_{n_2}$, where $n_2=m_2$ or
$n_2=n_1-m_2$, depending upon the parity of $r_1$.  we also have
$\gcd(n_1,n_2)=1$ and thus $n_1=r_2n_2+n_3$, with $n_3<n_2$ and
$\gcd(n_2,n_3)=1$. Thus, from the Euclidean algorithm, we can generate
the sequence $n_1,n_2,\dots,n_{f-1},n_f$ with $n_{f-1}-n_f=1$.
Corresponding to such a sequence on the chain, we can generate the
operator pairs $(z_{p_k},z_{p_k'})$, with $z_{p_1}=z_{1}$,
$z_{p_1'}=z_{n_1}$, $z_{p_2}=z_{(r_2-1)n_2}$, and $z_{p_2'}=z_{r_2n_2}$,
with $p_2-p_2'=n_1$ and $p_1'-p_2'=n_3$. Then we can generate $z_{p_3}$,
with $p_3=p_2'-r_3n_3$ and $p_3'=p_2'-(r_3-1)n_3$. We have $p_3-p_2=n_4$
and $p_3-p_3'=n_3$.  We can then generate $p_4$ and $p_4'$ between $p_3$
and $p_3'$ with $p_3'-p_4'=n_3$ and $p_4'-p_4=n_4$. We repeat this
process until we get $p_f$ and $p_f'$ where one of the two is the
neighbor of $p_{f-1}$ or $p_{f-1}'$. Thus, we can generate two operators
$z_a$ and $z_b$ such that $a-b=1$, and by Result~\ref{result:diff}, we
derive the controllability.

If $\gcd(N+1,n_1)=g>1$, then analogous to the above argument, we can
similarly generate $n_2,n_3,\dots,n_f$ with $n_f=g$, and similarly
the pair $(z_{p_j},z_{p_j'})$, with $p_j-p_j'=n_j$, $j=1,\dots,f$,
and $n_1>n_2>\cdots>n_f=g$, and then we can generate $z_g$. Thus all
operators $z_\ell$ generated from $z_k$ and $H_0$ satisfy
$\ell=sg$, and the operators $z_m$ with $g\mid m$ can not be
individually generated, so the dynamical Lie algebra generated from
$z_k$ and $H_0$ is strictly smaller than $U(N)$, and the system is
not controllable.

\end{IEEEproof}

Next, we try to demonstrate the above constructive proof through
examples. For $N=11$ and $n_1=k=5$, we have
\begin{align*}
N+1&=12=2\times 5+2=r_1n_1+m_2\\
n_1&=5 =2\times 2+1=r_2n_2+n_3
\end{align*}
According to Results~\ref{result:nk} and~\ref{result:k-m}, we can
sequentially generate $z_5$, $z_{10}$, then $z_2$, $z_4$, and then
$z_1$.  Hence we have controllability by Theorem~\ref{thm:z1_control}.
However, if $N=14$ and $k=5$ then it is easy to see that the only
$z_{\ell}$ operators we can generate from $-iH_0$ and $z_5$, are $z_5$
and $z_10$, and the system is not controllable in this case.

It is worth noting that from the above proof, not only have we proved
the lack of symmetry is equivalent to controllability on $\H_1$ for a
uniform XX chain, but we have also derived the dynamical Lie algebra
$\LL$ when the system is not controllable.  Moreover, the proof implies
a result which is also useful when discussing the Heisenberg chain in
the following section:

\begin{result}
\label{result:coprime-neighbor}
Given that we can generate $z_{p_i}$, $i=1,2,3$, with $p_1<p_2<p_3$
and $\gcd(p_2-p_1,p_3-p_1)=1$, we can further generate two
neighboring $z_\ell$ and $z_{\ell+1}$ with $p_1\le \ell<\ell+1\le
p_3$.  Hence, the system is controllable.
\end{result}

\subsection{Heisenberg Chain}

Analogous to XX chain, the correspondence between the lack of symmetry
and the controllability is summarized in the following theorem:

\begin{theorem}
\label{thm:Heisen_controllability} For a Heisenberg chain of length $N$
with uniform coupling strengths and a local Z control $z_k$, the system
is controllable on $\H_1$ if and only if $\gcd(N,2k-1)=1$.
\end{theorem}

Again the condition for controllability is equivalent to the no-symmetry
condition.  First of all, we investigate the new operators generated by
$-iH_0$ and $-iH_1$. Assuming $N>2k$ and for $k>1$, similar to the XX
chain case, we have

\begin{align*}
[-iH_0,z_k]&\rightarrow y_{k-1,k}+y_{k+1,k}\\
[y_{k-1,k}+y_{k+1,k},z_k]&\rightarrow x_{k-1,k}+x_{k+1,k}\\
-iH_2 &\equiv-iH_0-(x_{k-1,k}+x_{k+1,k})\\
[-iH_2,x_{k-1,k}+x_{k+1,k}]&\rightarrow y_{k-2,k}+y_{k+2,k}\\
[y_{k-2,k}+y_{k+2,k},z_k]&\rightarrow x_{k-2,k}+x_{k+2,k}\\
\cdots\cdots &\cdots \cdots\cdots\\
[-iH_2,x_{2,k}+x_{2k-2,k}]&\rightarrow y_{1,k}+y_{2k-1,k}\\
[y_{1,k}+y_{2k-1,k},z_k]&\rightarrow x_{1,k}+x_{2k-1,k}\\
[-iH_2,x_{1,k}+x_{2k-1,k}]&\rightarrow y_{1,k}+y_{2k,k}\\
[y_{1,k}+y_{2k,k},z_k]&\rightarrow x_{1,k}+x_{2k,k}\\
[-iH_2,x_{1,k}+x_{2k,k}]&\rightarrow x_{2,k}+x_{2k+1,k}\\
\cdots\cdots &\cdots \cdots\cdots\\
[-iH_2,x_{k-2,k}+x_{3k-3,k}]&\rightarrow x_{k-1,k}+x_{3k-2,k}\\
[-iH_2,x_{k-1,k}+x_{3k-2,k}]&\rightarrow x_{3k-1,k}\\
[x_{3k-1,k},z_k]&\rightarrow y_{3k-1,k}\\
[x_{3k-1,k},y_{3k-1,k}]&\rightarrow z_{3k-1}
\end{align*}

Thus, from $z_k$, we can sequentially generate $x_{k-1,k}+x_{k+1,k}$,
$\cdots$, $x_{k-m,k}+x_{k+m,k}$, until $x_{1,k}+x_{2k-1,k}$. When we
generate some $x$ operator(e.g. $x_{1,k}+x_{2k-1,k}$), we can always
generate the corresponding $y$ operator(e.g. $y_{1,k}+y_{2k-1,k}$) and
vice versa, so in the following, we only concentrate on the $x_{i,j}$
operators that can be generated. Different from XX model, and due to the
component $z_1$ in $-iH_2$, the next operator we can generate is
$x_{1}+x_{2k,k}$, rather than $x_{2k,k}$ as in the XX chain case. Then
we can sequentially generate $x_{j,k}+x_{j+2k-1,k}$, and finally we get
$x_{k+(2k-1),k}$, and hence $z_{k+(2k-1)}$. Analogously, starting from
$z_{k+(2k-1)}$, we can sequentially generate
$x_{k+(2k-1)-m,k+(2k-1)}+x_{k+(2k-1)+m,k}$, $m=1,\dots,2k-2$, and
finally we get $x_{k+2(2k-1),k}$ and hence $z_{k+(2k-1)}$. Continuing
such process, we have

\begin{result}
\label{result:nk-Heisenberg}
For Heisenberg chain, from $z_k$ and $H_0$, we can sequentially
generate $z_{k+m(2k-1)}$, with $1\le m\le r$ and $k+r(2k-1)<N$. In
particular, if $N=k+r(2k-1)$, then the system is controllable.
\end{result}

Next, similar to the XX chain case, we have the following result

\begin{result}
\label{result:Heisenberg-neighbor}
For Heisenberg chain, if from $H_0$ and $H_1$ we can generate two
neighboring $z_{k_1}$ and $z_{k_2}$ with $k_1-k_2=1$, then the
system is controllable.
\end{result}

\begin{figure}
\begin{tabular}{c}
\centerline{\includegraphics[width=\columnwidth]{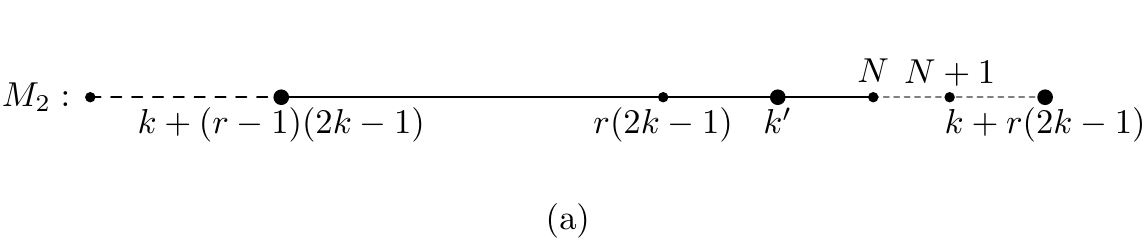}}\\
\centerline{\includegraphics[width=\columnwidth]{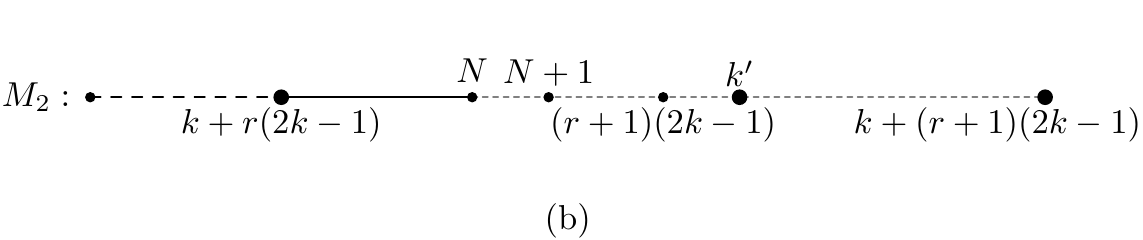}}
\end{tabular} \caption{Heisenberg chain $M_1$ is modified into the model $M_2$,
by extending the chain length to (a) $N'=k+r(2k-1)$, if
$r(2k-1)<N<k+r(2k-1)$; (b) $N'=k+(r+1)(2k-1)$, if $k+r(2k-1)\ge
N<(r+1)(2k-1)$. In both cases, we can generate $z_{k'}$ with
$k'=2N+1-(k+r(2k-1))$.} \label{fig2}
\end{figure}

Now we are ready to prove Theorem~\ref{thm:Heisen_controllability}.
Assuming $r(2k-1)<N<(r+1)(2k-1)$, for convenience of analysis, we
construct a modified model $M_2$ (FIG.~\ref{fig2}), corresponding
to the original Heisenberg chain,$M_1$ in~(\ref{eqn:XXZ-H0}).

If $r(2k-1)<N<k+r(2k-1)$, then in $M_2$, we extend the original
chain to length $N'=k+r(2k-1)$, and add new XX couplings in $H_0$ between the
nearest neighbors $x_{j,j+1}$, $N\le j\le N'-1$, but delete the term
$z_N$. Thus, in the modified system Hamiltonian $H_0$ the right end of $M_2$ is
of XX-type interaction,
rather than Heisenberg-type. As a result, in the modified model, starting from
$z_k$, we
can generate $z_j$ with $j=k+(r-1)(2k-1)$, and from $z_j$, we can
generate $x_{2j-N,k}+x_{N,k}$, the same as in the original Heisenberg model.
Then continuing the calculation,
in the modified model $M_2$, we
then generate $x_{2j-N-1,j}+x_{N+1,j}$ (corresponding to
$x_{2j-N-1,j}+x_{N,j}$ in $M_1$), then $x_{2j-N-2,j}+x_{N+2,j}$, and
finally $x_{k+r(2k-1),j}$ and $z_{k+r(2k-1)}$, which corresponds to
the generated operator $z_{k'}$ in $M_1$, with
$k'=2N+1-(k+r(2k-1))>k$(FIG.~\ref{fig2} (a)). If we make the
following identification: node $N+1$ in $M_2$ corresponding to node
$N$ in $M_1$, and node $N+2$ in $M_2$ corresponding to node $N-1$ in
$M_1$, and so forth, then the operators generated from $z_j$ in
$M_2$ are 1-to-1 corresponding to the $z$ operators generated in
$M_1$. Moreover, we can see that all these corresponding pairs of
nodes are mirror symmetric with respect to the middle point between
nodes $N$ and $N+1$. Hence, once a new operator is generated in
$M_2$, we can always get the corresponding new operator generated in
$M_1$ through the mirror symmetry. For instance, if we can generate
two neighboring operators $z_{\ell}$ and $z_{\ell+1}$ in $M_2$, they
must correspond to two neighboring $z$ operators in $M_1$ as well.

If $k+r(2k-1)<N<(r+1)(2k-1)$, then in $M_2$ we extend the spin
length to $k+(r+1)(2k-1)$. Since we have generated $z_{k+r(2k-1)}$
in $M_1$, by mirror symmetry, it corresponds to the operator
$z_{k'}$ which can be generated in $M_2$, with
$k'=2N+1-(k+r(2k-1))>k$(FIG.~\ref{fig2} (b)).

To complete the proof of the main theorem, let $r(2k-1)\le
N<(r+1)(2k-1)$.  From the previous discussion, we only need to discuss
what new operators can be generated for the modified model $M_2$, and by
the 1-to-1 correspondence between $M_1$ and $M_2$, we can recover what
new operators can be generated for $M_1$.  In both of the two cases
$r(2k-1)<N<k+r(2k-1)$ and $k+r(2k-1)<N<(r+1)(2k-1)$, we can always
generate $z_{k'}$ with $k'=2N+1-(k+r(2k-1))>k$, satisfying
\begin{align*}
   k'-k=2N-(r+1)(2k-1).
\end{align*}

If $\gcd(N,2k-1)=1$, we have $\gcd(k'-k,2k-1)=1$. If
$r(2k-1)<N<(r+1)(2k-1)$, we have $p_1\equiv
k+(r-1)(2k-1)<k'<k+r(2k-1)\equiv p_3$; if $k+r(2k-1)<N<(r+1)(2k-1)$,
we have $p_1\equiv k+r(2k-1)<k'<k+(r+1)(2k-1)\equiv p_3$. But in
both cases, $p_3-p_1=2k-1$ and $(p_3-p_1,k'-p_1)=1$. Hence, the
conditions of Result~\ref{result:coprime-neighbor} are satisfied,
and we can generate two neighboring operators $z_{\ell'}$ and
$z_{\ell'+1}$ with $p_1\le \ell'<\ell'+1\le p_3$ in $M_2$, which
must correspond to two neighboring $z$ operators in $M_1$. Hence the
original system $M_1$ is controllable.

If $\gcd(N,2k-1)=g>1$, then by the same reasoning for XX chain, the only
$z_j$ operator we can generate are $z_j=z_{mg}$, i.e., $j$ is multiple
of $g$. And all the other $z_j$ operator are coupled with other
operators. Hence we cannot generate all $z_j$ and the dynamical Lie
algebra is strictly smaller than $U(N)$.

The same arguments can be applied to show that lack of symmetry is also
the necessary and sufficient condition for controllability on $\H_1$ for
for $\kappa=-1$, and other cases could be studied similarly.  So far we
have fully characterized the 1-to-1 correspondence between the external
symmetries of $H_0$ and $H_1$, and the non-controllable cases of the
system.

\section{Branched Networks}
\label{section:networks}

In previous sections we have studied symmetries and controllability for
the simplest type of spin network, i.e., the linear chain.  The same
techniques can be applied to more complex spin networks.

\begin{figure}
\begin{tabular}{c}
\centerline{\includegraphics[width=\columnwidth]{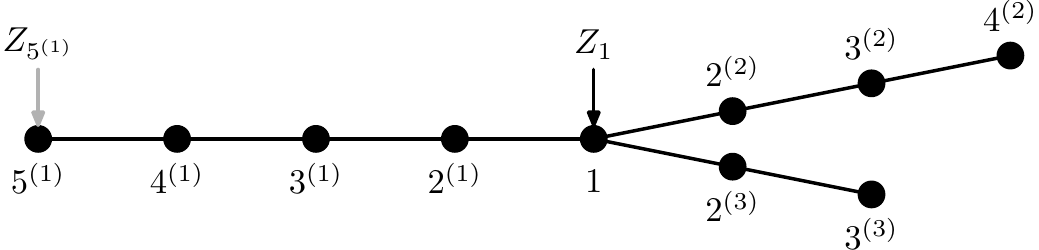}}\\
\end{tabular} \caption{Star shape spin networks with three subchains,
satisfying subchain lengths are pairwise coprime. We find that in
both cases (i) $H_1=z_1$, and (ii) $H_1=z_{5^{(1)}}$ the system is
controllable. }  \label{fig3}
\end{figure}

Besides the spin chain, the next simplest model of a spin network is the
star shape branched network, with a central spin, connected with $m$
number of spin subchains.  For $m=2$ this is equivalent to the chain
case.  The first non-trivial case is $m=3$, i.e., the T-shape spin
network, as illustrated in FIG.~\ref{fig3}. For a uniformly-coupled
XX-type branched network, the central spin can be indexed as $1$ and the
$m$ number of branches indexed as $2^{(p)}, \dots, \ell_p^{(p)}$,
$p=1,\ldots,m$. On $\H_1$, the branched network Hamiltonian and the
local control Z Hamiltonian on node $k^{(q)}$ are written as:
\begin{align*}
H_0&=\sum_{p=1}^{m}\Big(
x_{1,2^{(p)}}+\sum_{j^{(p)}}x_{j^{(p)},{j+1}^{(p)}}\Big)\\
H_1&=z_{k^{(q)}}
\end{align*}

We start with the simplest symmetry to identify, permutation symmetry.
If the local control is on one of the three subchains, say,
$H_1=z_{k^{(1)}}$, and the other two branches are of the same length,
$\ell_2^{(2)}= \ell_3^{(3)}$, then there is a permutation symmetry, and
the system is not controllable. In the following, we shall discuss cases
of the position of spin $k^{(q)}$.

\textbf{Case 1}: We control the far end of one branch(for example,
$H_1=Z_{5^{(1)}}$ in FIG.~\ref{fig3}).

We assume $H_1=z_{\ell_1^{(1)}}$. Analogous to the discussion of XX
chain, we can sequentially generate
$z_{{\ell_1-1}^{(1)}},\dots,z_{{2}^{(1)}},z_1$ as well as the associated
$x_{j,j+1}$ and $y_{j,j+1}$. Thus we can generate $z_1$ and the XX chain
Hamiltonian formed by the other two subchains.


According to Theorem~\ref{thm:XX}, we have if
$\gcd(\ell_2^{(2)},\ell_3^{(3)})=1$, then the system is controllable;
otherwise, it is not controllable.

\textbf{Case 2}: We control the central spin $z_1$ (for example,
$H_1=Z_{1}$ in FIG.~\ref{fig3}).

If two of the subchains are the same length, then the system is
not controllable. If the chains are of different length, then
motivated by the results in Theorem~\ref{thm:XX}, we have the
following conjecture:

\begin{conjecture}\label{conj:branch}
For star shape XX-type networks with three subchains, and local Z
control on the central spin $1$, there is no external symmetry and
the system is controllable if and only
$\gcd(\ell_j^{(j)},\ell_k^{(k)})= 1$ for $j\ne k$.
\end{conjecture}

For small values of $N$, we have calculated the symmetry and the dynamical Lie
algebra $\LL$ for all different patterns of such branched network, with a few
examples illustrated in Table~\ref{XX-branch}, all satisfying
Conjecture~\ref{conj:branch}.

\begin{table}
\begin{tabular}{ccccc}
N & $(\ell_1^{(1)},\ell_2^{(2)},\ldots,\ell_m^{(m)})$ & Symmetry & $\dim(\LL)$
&
Controllable \\\hline
10 & (5,4,3)                                   & No               & 100
& Yes \\
11 & (6,4,3)                                   & Yes               & 65
& No \\
12 & (6,5,3)                                   & Yes               & 101
& No \\
12 & (7,4,3)                                   & No               & 144
& Yes \\
13 & (6,5,4)                                   & Yes               & 144
& No \\
14 & (7,5,3,2)                                 & No               & 196
& Yes
\end{tabular}
\caption{Symmetry and controllability in $\H_1$ for different XX
branched networks.}
\label{XX-branch}
\end{table}

For Heisenberg-type of branched networks, we can similarly calculate the
symmetry and $\LL$ for controllability, with a few examples illustrated
in Table~\ref{Heisen-branch}. However, the explicit relationship between
the values of the parameters
$(N,\ell_1^{(1)},\ell_2^{(2)},\ldots,\ell_m^{(m)})$ and the patterns of
branched networks with symmetry requires further investigation in the
future.

\begin{table}
\begin{tabular}{ccccc}
N & $(\ell_1^{(1)},\ell_2^{(2)},\ldots,\ell_m^{(m)})$ & Symmetry & $\dim(\LL)$
&
Controllable \\\hline
8 & (2,3,5) & Yes & 50 & No\\
9 & (2,4,5) & Yes & 65 & No\\
9 & (2,3,6) & No & $9^2$ & Yes\\
10 & (3,4,5) & Yes & 65 & No\\
10 & (2,4,6) & No & $10^2$ & Yes\\
10 & (2,3,7) & No & $10^2$ & Yes\\
11 & (3,4,6) & No & $11^2$ & Yes\\
12 & (2,4,8) & Yes & 65 & No\\
12 & (3,5,6) & No & $12^2$ & Yes\\
13 & (4,5,6) & No & $13^2$ & Yes\\
14 & (4,5,7) & No & $14^2$ & Yes\\
14 & (2,4,10) & No & $14^2$ & Yes\\
15 & (2,4,11) & Yes & 122 & No
\end{tabular}
\caption{Symmetries and controllability for Heisenberg branched networks in $\H_1$.}
\label{Heisen-branch}
\end{table}

\section{Conclusion}
\label{section:conclusion}

In this work we have studied symmetries and controllability of XXZ spin
networks subject to local Z-controls on the single excitation subspace.
External symmetries for such systems can be easily characterized: such
symmetries exist if and only if there are eigenstates of the system
Hamiltonian that have no overlap with the control node.  If no such
symmetries exist then the system is indecomposable.  Unlike systems where
we control the coupling between two spins (or an edge in the associated
graph of the spin network), indecomposable systems which have
$Z$-controls applied to one or more spins have internal Lie algebra
symmetries only in very exceptional cases.

For linear XXZ chains we have further characterized all possible values
of $\kappa$ that allow the system to have external symmetries on the
single excitation subspace.  We find that there are at most countably
many values of $\kappa$ which permit any external symmetries, i.e.~for a
generic $\kappa$ there will be no symmetries, and we expect the system
to be controllable for any local Z control.  However, for the values of
$\kappa$ that are most relevant in real physical systems: $\kappa=0$ for
XX coupling, $\kappa=1$ for Heisenberg coupling and $\kappa=-1$ for
dipole coupling, there are symmetries in many cases.  The existence or
absence of symmetries depends on the position of the controlled node in
the chain relative to the length of the chain and the type of coupling
in a very peculiar manner.  This shows that the choice of controlled
node --- or the placement of the ``actuator'' in control terminology ---
is very significant.

For chains with non-uniform coupling strengths, we find that the
inhomogeneity usually breaks symmetries which are present for the
uniformly-coupled case with the same topology, as one might
expect. Surprisingly however, there are certain symmetries which are
robust even in the presence of inhomogeneities.  Absence of symmetries
is a necessary condition for controllability.  In many cases it also
appears to be a sufficient condition but there are examples of systems
without symmetries that are not controllable.  We have shown that for
uniform XX and Heisenberg chains, lack of symmetry is the necessary and
sufficient condition for the system's controllability on $\H_1$.
Finally, we have shown how to apply the techniques used to establish
controllabity for spin chains to more complex networks such as branched
networks.  We have characterized the possible symmetries and propose a
conjecture of the controllability condition.

Similar techniques could be applied to discuss the more complex spin
networks, as well as the relationship between symmetry, controllability
and actuator placement in other excitation subspaces.  For instance, it
could be applied to explain the observation that an antiferromagnetic
chain with local end-spin control appears to be controllable in the
largest excitation subspace~\cite{WSBB}.

\section*{Acknowledgments}
We thank Abolfazl Bayat, Sougato Bose, Alastair Kay, Francesco Buscemi,
Nilanjana Datta, Baojiu Li and Tony Short for valuable discussions, and
acknowledge funding from EPSRC ARF Grant EP/D07192X/1 and Hitachi, the
Cambridge Overseas Trust, Hughes Hall and the Cambridge Philosophical
Society for support.


\end{document}